\documentclass[twocolumn,preprintnumbers,amsmath,amssymb]{revtex4}

\usepackage{graphicx}
\usepackage{dcolumn}
\usepackage{bm}
\usepackage{hyperref}
\usepackage{amsmath}
\usepackage{epstopdf}
\usepackage{color}
\usepackage{verbatim}
\usepackage[normalem]{ulem}

\definecolor{LightGrey}{RGB}{211, 211, 211}

\begin{document}

\title{Shaping the spectrum of terahertz photoconductive antenna by frequency-dependent impedance modulation}

\author{\bf D. V. Lavrukhin$^{1,2}$,
A. E. Yachmenev$^{1,2}$,
A. Yu. Pavlov$^{1,2}$,
\mbox{R. A. Khabibullin$^{1,2,5}$},
Yu. G. Goncharov$^{2}$,
I. E. Spektor$^{2}$,
G. A. Komandin$^{2}$,
S. O. Yurchenko$^{3}$,
N. V. Chernomyrdin$^{2,3,4}$,
K. I. Zaytsev$^{2,3,4}$,
D. S. Ponomarev $^{1,2,5}$
}

\affiliation
{$^1$\mbox{Institute of Ultra High Frequency Semiconductor Electronics of RAS, Moscow 117105, Russia}\\
$^2$\mbox{Prokhorov General Physics Institute of RAS, Moscow 119991, Russia}\\
$^3$\mbox{Bauman Moscow State Technical University, Moscow 105005, Russia}\\
$^4$\mbox{Sechenov First Moscow State Medical University, Moscow 119991, Russia}\\
$^5$\mbox{Center for Photonics and 2D Material, Moscow Institute of Physics and Technology,Dolgoprudny 141700, Russia}\\
}


\begin{abstract}
    In this paper, we report on an approach
    for shaping the spectra of THz pulse generation
    in photoconductive antennas (PCAs)
    by frequency-dependent impedance modulation.
    We introduce a theoretical model
    describing the THz pulse generation in PCAs
    and accounting for impedances of the photoconductor and of the antenna.
    In order to showcase an impact of frequency-dependent impedance modulation on the spectra of THz pulse generation,
    we applied this model to simulating broadband PCAs with log-spiral topology.
    Finally, we fabricated two different log-spiral PCAs and characterized them experimentally using the THz pulsed spectroscopy.
    The observed results demonstrate excellent agreement between the theoretical model and experiment,
    justifying a potential of shaping the spectra of THz pulse generation in PCA
    by modulation of frequency-dependent impedances.
    This approach makes possible optimizing the PCA performance
    and thus accommodating the needs of THz pulsed spectroscopy and imaging
    in fundamental and applied branches of THz science and technologies.
\end{abstract}

\maketitle

\section*{Introduction}
\label{CHAPTER:Introduction}

Since the first observations of terahertz (THz) radiation
\cite{PhysRevSerI.4.4.314.1897},
it attracts considerable attention
due to the specificity of THz wave -- matter interaction:
the THz dielectric response
reveals low-energy molecular vibrations and structural features of the media
\cite{YunShikLeeBook2009}.
This allows to utilize
THz spectroscopy and imaging
for solving numerous fundamental and applied problems
in condensed matter physics
\cite{RevModPhys.83.2.543.2011}
and material science
\cite{OptExp.16.21.17039.2008},
gas sensing
\cite{OptLett.21.24.2011.1996}
and chemistry
\cite{JPharmPharm.59.2.209.2007},
biology and medicine
\cite{TrendsBiotech.34.10.810.2016}.
Despite considerable progress in THz technology,
THz instruments remain rare and expensive;
thus, stimulating the development of THz component base
\cite{IEEETransTerSciTech.1.1.133.2011,JPhysDApplPhys.45.30.303001.2012,JOpt.16.9.094007.2014,IEEETransTerSciTech.6.4.576.2016,ApplPhysLett.110.22.221109.2017,APLPhot.2.5.056.2017,BWO.Generators.2013,JPhysDApplPhys.51.13.135101.2018}.

Major progress in THz technology is a result of seminal research on semiconductor photoconductivity by Auston
\cite{ApplPhysLett.26.3.101.1975} and
is associated with the development of novel techniques
for generation and coherent detection of sub-picosecond THz pulses
\cite{ApplPhysLett.43.7.631.1983},
and related methods of THz pulsed spectroscopy
\cite{ApplPhysLett.55.4.337.1989}
and imaging
\cite{OptLett.20.16.1716.1995}.

During the past decades,
numerous approaches for generation and detection of
THz pulses with the use of
femtosecond laser radiation
have been developed,relying on various physical principles
and exploiting novel materials
\cite{YunShikLeeBook2009}.
Here, we can broadly identify the following trends:
\begin{itemize}

\item~\textit{THz pulse generation and detection
in accelerating charge carriers of semiconductors},
such as
low-temperature grown semiconductors (LT-GaAs or LT-InGaAs) and related heterostructures
\cite{ApplPhysLett.71.19.2743.1997,IEEETransTerSciTech.2.6.617.2012,TechPhysLett.43.11.1020.2017},
radiation-damaged semiconductors (GaAs or Si)
\cite{ApplPhysLett.54.24.2424.1989},
\cite{ApplOpt.36.30.7853.1997},
in ZnSe
\cite{JApplPhys.116.4.043107.2014},
GaAsBi
\cite{ApplPhysExp.5.2.022601.2012},
and others,
using the photoconductivity
\cite{ApplPhysLett.26.3.101.1975},
the photo-Dember effect
\cite{OptExp.18.5.4939.2010}
or the built-in electric field
\cite{OptExp.15.8.5120.2007}.

\item~\textit{THz pulse generation and detection in nonlinear media},
such as semiconductors (ZnTe, GaAs, GaP, InP, GaSe)
\cite{ApplPhysLett.85.18.3974.2004},
inorganic electrooptical crystals
(LiNbO$_{3}$, LiTaO$_{3}$)
\cite{OptLett.42.9.1704.2017,ApplPhysLett.56.6.506.1990},
organic crystals
\cite{ApplPhysLett.61.26.3080.1992},
and VO$_{2}$-films undergoing metal-insulator phase transition
\cite{Optica.2.9.790.2015},
using optical rectification
\cite{PhysRevLett.28.14.897.1972}
and electrooptical effects
\cite{ApplPhysLett.74.11.1516.1999}.

\item~\textit{THz pulse generation and detection
in accelerating charge carriers in gas plasma} (He, N, air, etc.)
\cite{PhysRevLett.71.17.2725.1993}
induced and probed by either the single-
\cite{PhysRevLett.96.7.075005.2006}
or dual-color
\cite{OptLett.38.11.1906.2013}
femtosecond laser beams.
\end{itemize}
Among these principles of (and materials for) THz pulse generation and detection,
the LT-GaAs and LT-InGaAs-based photoconductive antennas (PCAs)
represent the most prevalent type of THz pulsed emitter and detector.
They are widely applied in THz spectroscopy and imaging
due to simplicity, flexibility and reliability
of the PCAs' design and their technical characteristics.

Nevertheless, a rapid progress in THz science and technology
pushes further development of THz PCAs
into a realm of improving the optical-to-THz conversion efficiency
and optimizing the spectrum of THz pulse for different applications
\cite{PhotRes.4.3.A36.2016,IEEEJQuantumElectron.41.5.717.2005,IEEETransTerSciTech.4.5.575.2014,OptExp.22.11.12982.2014,OptExp.22.14.16841.2014}.
Typically, a spectrum of THz pulse is modeling when a PCA's impedance is frequency-independent and thus an influence of both antenna and photoconductor impedances on a spectrum shape is considered. In present article, in order to take into account this fact, we propose an approach for shaping a spectrum of THz pulse in broadband PCA by modulating its frequency-dependent impedance and improve correctness of a PCA's numerical description.

To reveal an impact of frequency-dependent impedance modulation on the THz pulse spectrum,
we proposed a theoretical model,
which describes the THz generation in PCA
and takes into account the impedances.
Then, to showcase an impact of frequency-dependent impedance modulation on the THz pulse spectrum,
this model is applied for studying the THz generation in two different broadband PCAs of the log-spiral topology.
Finally, we fabricate these PCAs and characterize them experimentally: we use them as an emitter in a THz pulsed spectrometer
to measure the spectra of THz pulse generation.
By comparing the results of theoretical and experimental study,
we demonstrate excellent agreement between the model and the experiment,
justifying our approach for shaping the spectrum of THz pulsed generation in PCA
by modulating frequency-dependent impedances of photoconductor and antenna.
The results of this study open a new way to optimize the THz PCA performance
for accommodating the needs of rapidly-developing THz science and technology,
particularly, in condensed matter physics, material science, chemistry, biology and medicine.

\section{Modelling the THz pulse generation in PCA}

\begin{figure}[!b]
    \centering
    \includegraphics[width=0.9\columnwidth]{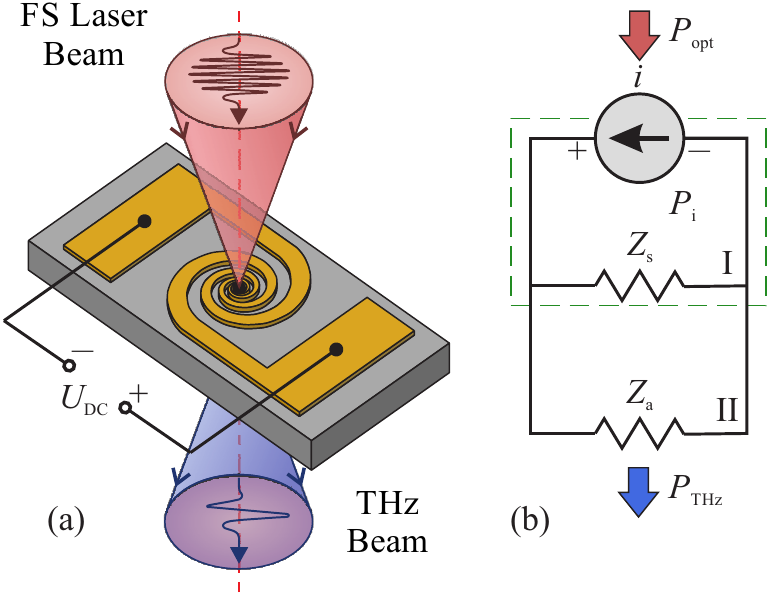}
    \caption{THz pulse generation in the LT-GaAs PCA:
    (a)~a scheme of the PCA;
    (b)~an equivalent circuit of the PCA.}
    \label{FIG:PCAs}
\end{figure}

We start from modelling the process of THz pulsed generation in PCA
and consider a low-temperature grown gallium arsenide (LT-GaAs) as a conventional material platform for PCA fabrication.
As shown in Fig~\ref{FIG:PCAs}~(a),
a femtosecond laser pulse irradiates the small area in the gap between electrodes in LT-GaAs photoconductor and results in generation of free carriers.
A DC electric field, applied to the electrodes, drives the motion of these free carriers,
resulting in the sub-picosecond-duration photocurrent and thus, the THz pulse generation.
The transient process occurs not only in photoconductor but in entire PCA and is described by Maxwell equations with appropriate boundary condition on metal and photoconductor's surface. Different antenna topology leads to different transient current distribution, traditionally characterized by antenna impedance value, that consequently results in spectral features of emitted THz waves.

Figure~\ref{FIG:PCAs}~(b) shows an equivalent circuit
\cite{IEEETransAntennasPropag.61.4.1538.2013,JIMTW.33.12.1182.2012},
applied for the PCA modelling.
According to the circuit,
the photoconductive gap, illuminated by the femtosecond pulse,
is considered as a source of photocurrent $i \left( t \right)$ (see Fig.~\ref{FIG:PCAs}~(b), part I),
for which we analytically calculate
an internal effective resistance $R_\mathrm{s}$
(i.e. a real part of an impedance $Z_\mathrm{s}$)
and a photocurrent power in Fourier-domain $P_\mathrm{i} \left( \nu \right)$
(here, $t$ and $\nu$ stand for the time and the frequency).
Then, the antenna impedance $Z_\mathrm{a}$
is estimated using numerical simulations,
and the electromagnetic coupling between a photoconductor and antenna (i.e. impedance matching factor)
is applied for estimating the emitted THz power spectrum $P_\mathrm{THz} \left( \nu \right)$ (see Fig.~\ref{FIG:PCAs}~(b), part II).
This modelling approach gives a non-self-consistent solution
for the THz pulse generation problem:
i.e. the photoconductive gap
as a source of photocurrent
and the log-spiral antenna arms (suppress certain frequency-domain modes and shapes the THz pulse spectrum)
as an external resistive load, are considered separately.

\subsection{Modelling the PCA Photocurrent}

In order to calculate the internal impedance $Z_\mathrm{s}$
and the power spectrum $P_\mathrm{i}$ of the current source,
we introduce a model of the time-dependent photocurrent of the PCA $i \left( t \right)$
in case of its irradiation with the ultra-short laser pulses.

Consider the PCA gap illuminated by the femtosecond laser pulse
with the Gaussian profile of instantaneous intensity\cite{JApplPhys.115.19.193105.2014}

\begin{equation}
   I_\mathrm{opt} \left( t \right) = \frac{ P_\mathrm{opt} \eta_\mathrm{opt}}{ \sqrt{ \pi } f_\mathrm{rep} \tau_\mathrm{p} }
                                     \exp \left( - t^2 / \tau_\mathrm{p}^2 \right),
      \quad
      \tau_\mathrm{p}= \frac{ \tau_\mathrm{opt} }{ 2 \sqrt{ \ln 2 } },\\
   \label{equation:IL}
\end{equation}
where
$P_\mathrm{opt}$ is an average laser power,
$\eta_\mathrm{opt}$ is a fraction of laser power which is absorbed in photoconductor,
$f_\mathrm{rep}$ is a pulse repetition rate,
$\tau_\mathrm{p}$ and $\tau_\mathrm{opt}$ define
a duration of the Gaussian pulse
either as a parameter of the Gaussian function or as a full-width at half-maximum (FWHM).
Assuming the uniform character of the antenna gap irradiation,
the carrier transport in photoconductor
can be described in the framework of 1D-Drude-Lorentz model
\cite{YunShikLeeBook2009}
with a cross-section area $\Delta S$.
Since the effective mass of electrons in LT-GaAs is much smaller than that of holes,
the electrons are considered as dominant charge carriers.
We use the following expression for the time-dependent non-equilibrium concentration of electrons in the PCA's gap
$n_\mathrm{e} \left( t \right)$, i.e. a response on the excitation by the electromagnetic wave
featuring the Dirac-function-like instantaneous intensity $\delta \left( t \right)$ with energy $\Delta E$
\cite{YunShikLeeBook2009}

\begin{equation}
   \begin{gathered}
      n_\mathrm{e} \left( t \right) = \frac{ \lambda_\mathrm{opt} }{ h c l } \frac{\Delta E }{\Delta S}\eta_\mathrm{opt}
                                      \exp \left( \frac{- t } { \tau_\mathrm{c} } \right),\\
      \eta_\mathrm{opt}=
      \bigg[1-
              \bigg( \frac{n_\mathrm{GaAs}-1}{n_\mathrm{GaAs}+1}\bigg)^2 \bigg]
      \cdot
              \bigg(1-exp(-\alpha d)\bigg),
   \end{gathered}
   \label{equation:ne}
\end{equation}
where
$\tau_\mathrm{c}$ is a carrier lifetime
($\tau_\mathrm{c} \gg \tau_\mathrm{p}$),
$l$ is a length of the illuminated gap,
$h \simeq 6.63 \times 10^{-34}$~Joule$\cdot$sec is the Planck constant,
$c \simeq 3 \times 10^{8}$~m/sec is the speed of light in
free space,
$\lambda_\mathrm{opt}$ is a central wavelength of the femtosecond laser radiation,
$n_\mathrm{GaAs}$ and $\alpha$ are refractive index and
absorption coefficient for LT-GaAs layer (of thickness $d$) respectively.

In our consideration, the electron density under illumination satisfies the condition $\omega_\mathrm{p}\tau_\mathrm{s}\sim1$,
where \mbox{$\omega_\mathrm{p}^{2} = n_\mathrm{e} e^2 / (\varepsilon_\mathrm{0} m_\mathrm{e} )$} stands for a plasma frequency,
$\tau_\mathrm{s}$ is an electron momentum relaxation time ($\tau_\mathrm{s} \ll \tau_\mathrm{c}$),
$\varepsilon_\mathrm{0}$ and $m_\mathrm{e}$ are the
dielectric constant of the medium
and the electron effective mass, respectively. This condition
allows to neglect the screening of the external bias electric field $E_\mathrm{DC}$,
which originates from separation of electron-hole pairs~\cite{JOSAB.13.11.2424.1996}.
We assume the average electron velocity $v_\mathrm{e} \left( t \right)$
according to the following equation~\cite{YunShikLeeBook2009}

\begin{equation}
   v_\mathrm{e} \left( t \right) = \mu_\mathrm{e} E_\mathrm{DC} \big( 1 - \exp \left( - t / \tau_\mathrm{s} \right) \big),\\
   \label{equation:ve}
\end{equation}
where
$\mu_\mathrm{e} = e \tau_\mathrm{s} / m_\mathrm{e}$ is an electron mobility.

The photocurrent can be expressed
as a convolution of the instantaneous intensity of laser pulse  $I_\mathrm{opt} \left( t \right)$ (Eq.~\ref{equation:IL})
with the responses to the electron concentration $n_\mathrm{e} \left( t \right) $ (Eq.~\ref{equation:ne})
and with the electron velocity $v_\mathrm{e} \left( t \right)$ (Eq.~\ref{equation:ve})
\cite{IEEEJSelTopQuantElectron.7.4.615.2001}:

\begin{equation}
        i \left( t \right) = e \int \limits_{0}^{\infty} I_\mathrm{opt} \left( t - t' \right) \big[ n_\mathrm{e} \left( t' \right) v_\mathrm{e} \left( t' \right) \big] dt',
        \label{EQ:ConvolutionIntegral}
\end{equation}
where $e \simeq 1.6 \times 10^{-19}$~C is the elementary charge.
Integration of \eqref{EQ:ConvolutionIntegral} leads to the analytical expression for the photocurrent

\begin{equation}
   \begin{gathered}
        i \left( t \right) = \frac{U_{\mathrm{DC}} }{B}
                                    \bigg(
                                    \exp          \bigg(   \frac{ \tau_\mathrm{p}^{2} } { 4 \tau_\mathrm{c}^{2} }
                                                         - \frac{ t                   } { \tau_\mathrm{c}       }
                                                  \bigg)
                                    \mathrm{erfc} \bigg(   \frac{ \tau_\mathrm{p}     } { 2 \tau_\mathrm{c}     }
                                                         - \frac{ t                   } { \tau_\mathrm{p}}
                                                  \bigg)
                                    -\\
                                    \exp          \bigg(   \frac{ \tau_\mathrm{p}^{2} } { 4\tau_{\mathrm{cs}}^2 }
                                                         - \frac{ t                   } { \tau_\mathrm{cs}}\bigg)
                                    \mathrm{erfc} \bigg(   \frac{ \tau_\mathrm{p}     } { 2\tau_\mathrm{cs}}
                                                         - \frac{ t                   } { \tau_\mathrm{p}}
                                                  \bigg)
                                \bigg),
    \end{gathered}
    \label{equation:Ipc}
\end{equation}
where
\begin{equation}
   \begin{gathered}
        \tau_\mathrm{cs}^{-1}  = \tau_\mathrm{c}^{-1}+\tau_\mathrm{s}^{-1}\simeq  \tau_\mathrm{s}^{-1},\\
        \mathrm{erfc} \left( x \right) = \frac{ 2 } { \sqrt{\pi}} \int \limits_{x}^{\infty} \exp \left( -t^{2} \right) dt,\\
        \frac{1}{B} = \frac{ P_\mathrm{opt}  \eta_\mathrm{opt}}{ f_\mathrm{rep}  }
              \frac{ \lambda_\mathrm{opt} }{ h c}
              \frac{e \mu_\mathrm{e}}{2 l^2},
   \end{gathered}
\end{equation}
and $U_\mathrm{DC}$ stands for the bias voltage
\begin{equation}
    U_\mathrm{DC}=E_\mathrm{DC} l.
\end{equation}

\begin{table}[!b]
    \centering
    \caption{Parameters of the photocurrent model}
    \includegraphics[width=0.9\columnwidth]{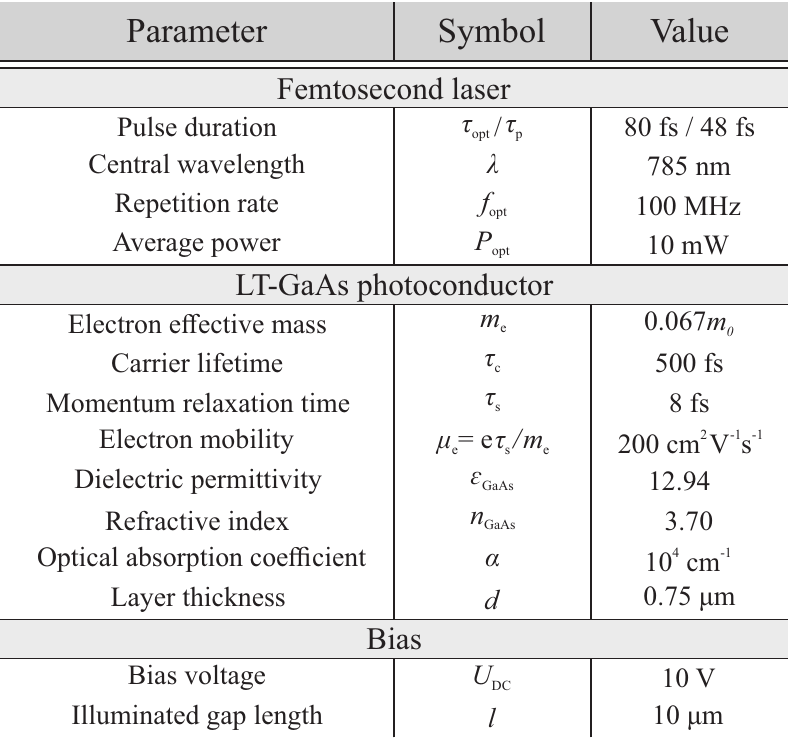}
    \label{tabular:ModelPAR}
\end{table}

From Eq.~\eqref{equation:Ipc},
we observe that the gap resistance has a time-dependent character. However, for simplicity,
we introduce an effective resistance of the photoconductive gap

\begin{equation}
   \begin{gathered}
       R_\mathrm{s} = \frac{ U_\mathrm{DC} \int \limits_{-\infty}^{+\infty} i \left( t \right) dt }
                           { \int \limits_{-\infty}^{+\infty} i^2 \left( t \right) dt }
       \simeq \\
       B \frac{ \tau_\mathrm{c} }{ \tau_\mathrm{c} - \tau_\mathrm{cs} }
       \exp \left( - \frac{ \tau_\mathrm{p}^2 } { 2 \tau_\mathrm{c}^2 } \right)
       \mathrm{erfc} \left( \frac{ \tau_\mathrm{p} } { \sqrt{2} \tau_\mathrm{c} } \right)
       \simeq B,
    \end{gathered}
    \label{equation:Rpc}
\end{equation}
and the Fourier-domain power of the photocurrent
\begin{equation}
    P_\mathrm{i} \approx 2  | \widetilde{i} \left( \nu \right) |^2 R_\mathrm{s} f_\mathrm{rep},
    \label{equation:Pe}
\end{equation}
where
\begin{equation}
    \begin{gathered}
       \widetilde{i}\left( \nu \right) \approx \frac{2 U_{\mathrm{DC}} } { B } \frac{\tau_\mathrm{c} - \tau_\mathrm{sc} } { 1 + j 2 \pi \tau_\mathrm{c} \nu} exp(- \pi^2 \tau_\mathrm{p}^2 \nu^2)
    \end{gathered}
    \label{EQ:FourierSpectrumOfPhotocurrent}
\end{equation}
stands for the Fourier spectrum of the photocurrent.

\begin{figure}[!t]
    \centering
    \includegraphics[width=0.9\columnwidth]{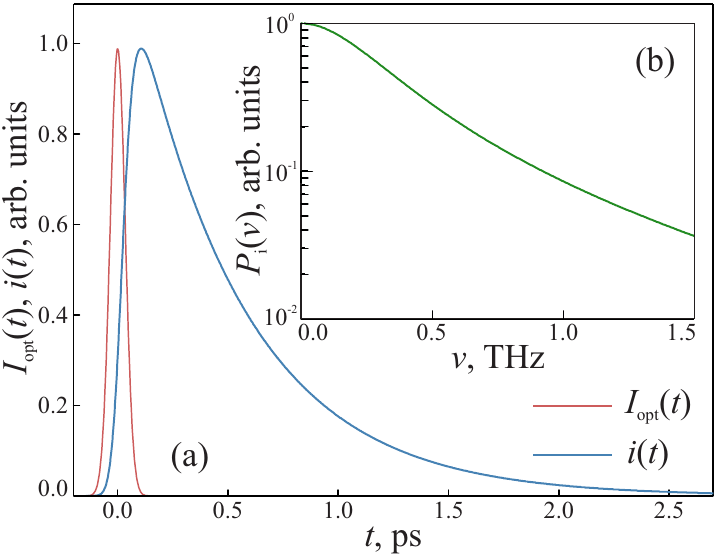}
    \caption{Results of modelling the photocurrent of LT-GaAs-based PCA:
    (a) a time-dependent photocurrent $i\left( t \right)$ (Eq.~\eqref{equation:Ipc});
    (b) a Fourier spectrum of photocurrent power $P_\mathrm{i} \left( \nu \right)$
    of the LT-GaAs PCA excited by the femtosecond laser pulse featuring
    (a) the instantaneous intensity $I_\mathrm{opt} \left( t \right)$ (Eq.~\eqref{equation:IL}). All values are normalized to maximum.}
    \label{FIG:Ipc}
\end{figure}

We computed the performance of the LT-GaAs PCA using the described model
Eqs.~\eqref{equation:IL}--\eqref{EQ:FourierSpectrumOfPhotocurrent}.
In these computations, we used the technical characteristics of the Toptica FemtoFErb780 laser,
as a source of the THz pulses, and the parameters of the LT-GaAs according to the Refs.
\cite{YunShikLeeBook2009, IEEETransAntennasPropag.61.4.1538.2013, JApplPhys.90.1303.2001}
(see Tab.~\ref{tabular:ModelPAR}).
The results of the PCA photocurrent modelling are shown in Figure~\ref{FIG:Ipc}~ where one can see the normalized instantaneous intensity of the femtosecond laser pulse irradiating the antenna gap
and the time-domain normalized photocurrent (a),
the normalized power spectrum of the photocurrent (b).
According to the Eq.~\eqref{equation:Rpc},
we determined the effective resistance of the photoconductor gap $R_\mathrm{s} \simeq 0.4$~$k \Omega$.

From the Fig.~\ref{FIG:Ipc}, one can see that the observed photocurrent $i \left( t \right)$
has a form of the non-symmetric pulse,
the peak of which is slightly delayed relative to the peak of the femtosecond excitation
and the duration of which is predominantly determined by the longest time scale
-- i.e the lifetime of the LT-GaAs photocarriers $\tau_\mathrm{c}=0.5$~ps.
The photocurrent power spectrum $P_\mathrm{i}$
decays with an increase of electromagnetic frequency $\nu$,
and possesses no remarkable spectral features.
Thereby, the modelling of the photocurrent allows for determining
main parameters of the current source in the equivalent circuit
(see Fig.~\ref{FIG:PCAs}~(b)),
which forms the basis for its further analysis.

\subsection{Accounting the PCA Topology}

An essential feature of the log-spiral PCA
is a broadband spectrum of generated THz pulse
appeared as a result of the absence of the cut-off frequency,
which is usually defined by antenna arms.
In such extended topology there is a non-uniform transient current distribution on arms’ surface. Thereby for each of the emitted THz wavelength the active area of the PCA is given by a number of the spiral turns,
with length of the last turn close to this wavelength \cite{JApplPhys.109.6.061301.2011,BOOK.AntennaTheory.2005}: $\lambda_\mathrm{GaAs} = \lambda / \sqrt{\varepsilon_\mathrm{eff}}$ (where $\lambda$ and $\lambda_\mathrm{GaAs}$ are the electromagnetic wavelength in the free space and in LT-GaAs, respectively,
and $\varepsilon_\mathrm{eff} = \left( 1 + \varepsilon_\mathrm{GaAs} \right) / 2 \approx 6.97$ is the effective real part of the dielectric permittivity of LT-GaAs  to account for the operation at the semiconductor-air interface).

The log-spiral arms are characterized by their own impedance (see Fig.~\ref{FIG:PCAs})
-- a complex frequency-dependent function $Z_\mathrm{a} = R_\mathrm{a} + i X_\mathrm{a}$,
comprised of a real (a radiation resistance $R_\mathrm{a}$)
and an imaginary (a reactance $X_\mathrm{a}$) parts
\cite{JIMTW.33.12.1182.2012,IEEETransAntennasPropag.61.4.1538.2013}.
In order to provide an efficient conversion of the photocurrent power to the THz beam power,
the antenna impedance $Z_\mathrm{a}$ should be matched with the impedance of the current source (see Fig.~\ref{FIG:PCAs}),
which, in general case, is also represented by the complex frequency-dependent function $Z_\mathrm{s} = R_\mathrm{s} + i X_\mathrm{s}$.
Mismatching of these impedances leads to weakening of electromagnetic coupling between photoconductor and antenna, i.e. reflection of the current oscillations' power from antenna metallization, and a consequent reduction both a PCA's photocurrent and optical-to-THz conversion efficiency.
\cite{ProcSPIE.6194.61940G.2006}:
\begin{equation}
    P_\mathrm{THz} = \eta_\mathrm{m} P_\mathrm{i},
    \qquad
    \eta_\mathrm{m} = 1 - \left| \frac { Z_\mathrm{a} - Z_\mathrm{s} } { Z_\mathrm{a} + Z_\mathrm{s} } \right|^{2},
    \label{equation:Nm}
\end{equation}
where $P_\mathrm{THz}$ stands for the THz-wave power and $\eta_ \mathrm{m}$ is a parameter accounting the electromagnetic coupling efficiency.

The impedance of the photocurrent source $Z_\mathrm{s}$
involves the above mentioned internal ohmic resistance $R_\mathrm{s}$ and reactance of a capacitor $X_\mathrm{s}$,
which accounts for accumulation of the charge carrier in the illuminated antenna gap as well as the screening effects.
The applied external electric field causes a separation of the photoexcited electrons and holes.
Being separated, these charges produce their own counter-directed screening electric field ~\cite{ApplPhysLett.88.16.161117.2006}, that
leads to screening and restricts the current oscillations~\cite{JApplPhys.109.6.061301.2011,IEEETransAntennasPropag.61.4.1538.2013}.
However, due to the applied simplified Drude model (Eq.~\eqref{equation:ne}),
which excludes the dynamics of the holes,
the screening and the reactive part of the photocurrent source impedance are neglected
-- i.e. $X_\mathrm{s} = 0$ and $Z_\mathrm{s} = R_\mathrm{s}$.

The impedance of the ideal infinite self-complementary log-spiral antenna
on the semi-infinite LT-GaAs ~\cite{JApplPhys.109.6.061301.2011}
comprised of the valuable real and negligible imaginary parts: $Z_\mathrm{a0} = R_\mathrm{a0} \approx 60 \pi / \sqrt{\varepsilon_\mathrm{eff}} \approx 71~\Omega \ll R_\mathrm{s}$.
This fact dissatisfies the condition of impedance matching and thus decreases the THz wave generation power. However, the frequency-dependent impedance of real log-spiral PCA
can be varied near it's theoretical value $Z_\mathrm{a0}$ by altering the antenna topology, which results in opportunity for modulating an emitted THz spectrum (i.e. spectrum shaping) by altering the $\eta_ \mathrm{m}$ value. It should be noted that the proposed approach does not contradict with the well-known far-field dipole-generation formalism ~\cite{YunShikLeeBook2009, JOSAB.13.11.2424.1996}
but extends it to the case of PCA with extended arms which is possessing essentially a non-uniform surface photocurrent distribution. This is of particular importance when considering broadband PCAs.

We consider two PCAs with different topology with the edges formed by the four self-complementary log-spirals
\begin{equation}
    \begin{gathered}
        r_\mathrm{1} = \exp \left( \alpha \varphi \right),
        \qquad
        r_\mathrm{2} = M r_\mathrm{1},\\
        r_\mathrm{3} = \exp \left( a \left( \varphi - \pi \right) \right),
        \qquad
        r_\mathrm{4} = M r_\mathrm{3},
    \end{gathered}
    \label{equation:Spiral}
\end{equation}
where $\alpha$
is an expansion rate
and $M$ is an arm thickness parameter
-- see Tab.~\ref{TAB:Table2}.
In order to perform adequate comparison of the antennas,
all the PCAs have the gap length of $10$~$\mu$m.

\begin{table}[!b]
\centering
  \caption{Topology of the log-spiral PCAs under consideration}
  \includegraphics[width=0.9\columnwidth]{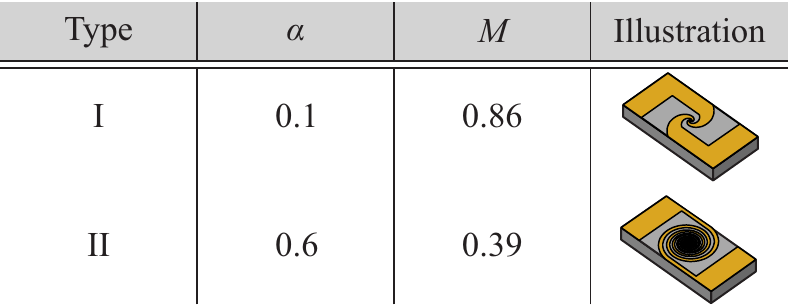}
  \label{TAB:Table2}
\end{table}

In order to estimate the impedance of these antennas,
we performed numerical simulations using the finite integration technique (FIT)
-- a convenient method of solving the Maxwell's equations
\cite{IntJNM:JNM240}.
An important feature of FIT is a possibility of the local mesh refinement
in the areas of either the strong field variations
or the tiny topological elements of the antenna.
The antenna excitation is simulated
by applying the common Gaussian-shaped time-dependent signal close to the gap.
For the PCAs featuring the described topology (see Tab.~\ref{TAB:Table2}),
the thickness of $0.35$~$\mu$m, and the infinitely-high conductivity
(perfect electric conductor),
we examine the complex currents and voltages originating in the PCAs,
as a result of this excitation.

We assume the antenna's electrodes to be deposited on the surface of the semi-infinite LT-GaAs,
which possesses the dielectric permittivity of $\varepsilon_\mathrm{GaAs} = 12.94$ (Tab.~\ref{tabular:ModelPAR})
and negligible dynamic conductivity. The dimensions of the simulations volume are $5 \times 3 \times 0.02$~mm$^3$
-- they include only the interface of the semiconductor with the antenna electrodes.
In order to prevent a non-physical backscattering
of the electromagnetic waves from the boundaries of the simulations space,
the latter is surrounded by $6$ perfectly matched layers.
In simulations we use the 3D tetrahedral mesh
with grid sizes in the range of $30$ to $0.3$~$\mu$m.

\begin{figure}[!t]
    \centering
    \includegraphics[width=0.9\columnwidth]{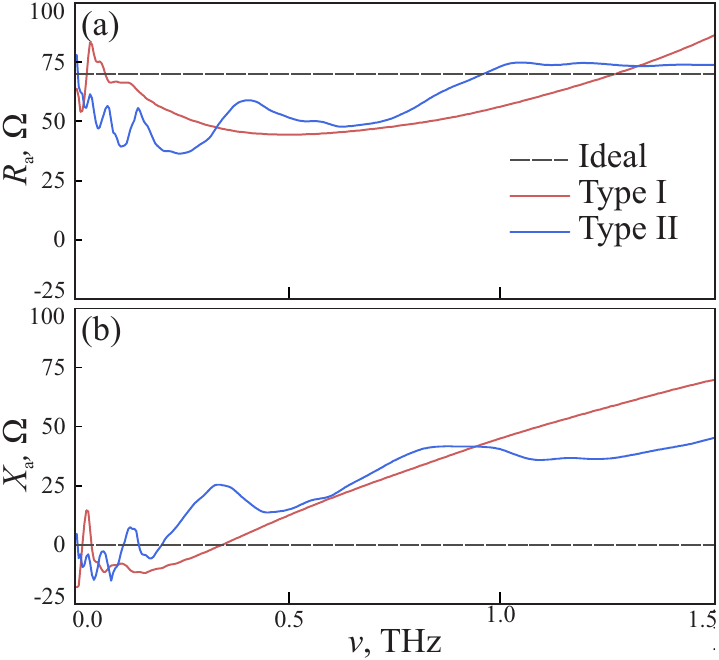}
    \caption{The results of modelling the THz pulse generation
    in log-spiral PCAs with two different topology:
    (a),~(b)~radiation resistance $R_\mathrm{a}$ and reactance $X_\mathrm{a}$ calculated for the two log-spiral PCAs (see Tab.~\ref{TAB:Table2})
    using the numerical simulations. The horizontal dashed lines
    represent the theoretical impedance $Z_\mathrm{a0} = R_\mathrm{a0} + i X_\mathrm{a0}$
    of the ideal infinite self-complementary log-spiral PCAs
    on the semi-infinite LT-GaAs.}
    \label{FIG:Za}
\end{figure}

Results of the numerical simulations
for the two antenna topologies (see Tab.~\ref{TAB:Table2}) are demonstrated in Fig.~\ref{FIG:Za}:
(a)~and (b)~show the radiation resistances $R_\mathrm{a}$ and reactances $X_\mathrm{a}$, respectively.
At high frequencies,
we limit the spectral range of numerical simulations by $1.5$~THz,
since, for the equal physical size of the simulation volume,
the numerical modelling at higher frequencies
requires lower spatial grid size and larger size of the mesh,
which significantly increases the length of calculations.
From the Fig.~\ref{FIG:Za}, we could notice that
the radiation resistance $R_\mathrm{a}$ varies near (primarily, below)
the radiation resistance of the ideal log-spiral antenna $R_\mathrm{a0}$,
shown by the dashed line,
while the reactance $X_\mathrm{a}$ features non-zero value (predominantly, positive)
and increases with frequency $\nu$.
It is important to note that the simulated PCAs demonstrate quite different behavior in their frequency-dependent impedances.

Considering Eq.~\eqref{equation:Nm},
we might expect that the obtained wavy-like structure of the PCA Type II impedance in Fig.~\ref{FIG:Za}
would lead to related spectral features in the emitted THz power spectrum ---
we will consider and discuss these features below together with the experimental data.

\section{Experimental Characterization of PCAs}

In order to test the results of theoretical modelling,
as well as to justify the impact of the frequency-dependent impedance modulation on the THz pulse spectrum,
we fabricated the LT-GaAs log-spiral PCAs (both Type I and II)
and experimentally characterized the THz pulse generation using LT-GaAs log-spiral PCAs as emitters in a THz pulse spectrometer.

\subsection{Fabrication of PCAs}

\begin{figure}[!t]
    \centering
    \includegraphics[width=0.9\columnwidth]{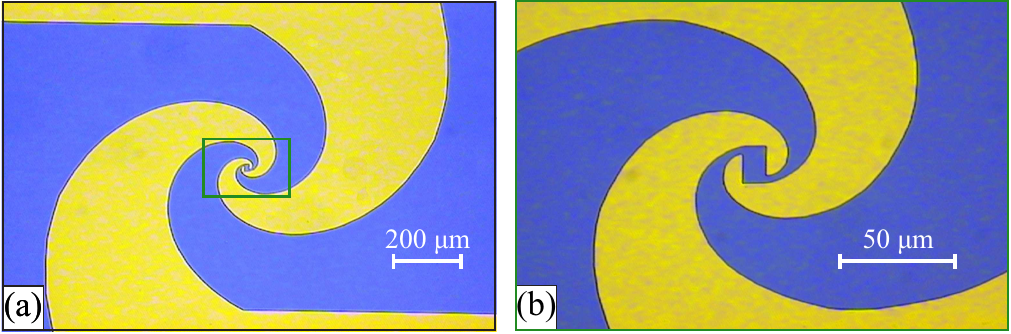}
    \caption{Microscopy of the log-spiral PCA Type I:
    (a)~a microscopic image of antenna's arms with its gap;
    (b)~a magnified image of the gap.}
    \label{FIG:AntennaMicroscopy}
\end{figure}

First, the $0.75$-$\mu$m-thick photoconductive LT-GaAs layer
was grown by the molecular-beam epitaxy at the temperature of $215^\circ$C,
which was followed by the \textit{in situ} annealing
at the temperature of $600^\circ$C during $20$~min~\cite{Semicond.51.4.509.2017}.
Second, the log-spiral antennas were formed on the surface of the LT-GaAs
using the contact UV-photolithography with the LOR5A-S1813 resist,
which is prepared in the tetramethylammonium~hydroxide solution in the absence of metal ions.
An oxygen plasma was used to remove residual resist,
and GaAs oxides are removed in the aqueous solution of the hydrochloric acid.
Finally, the antennas topology is formed
by the Ti-Pt-Au ($30$--$25$--$300$~nm) metallization lift-off.
All the technological aspects of the LT-GaAs PCA manufacturing
are described in details in Ref.~\cite{Semicond.51.9.1218.2017},
while the microscopic images of the PCA Type I are shown in Fig.~\ref{FIG:AntennaMicroscopy}.

\subsection{Experimental Setup}

\begin{figure}[!b]
    \centering
    \includegraphics[width=0.6\columnwidth]{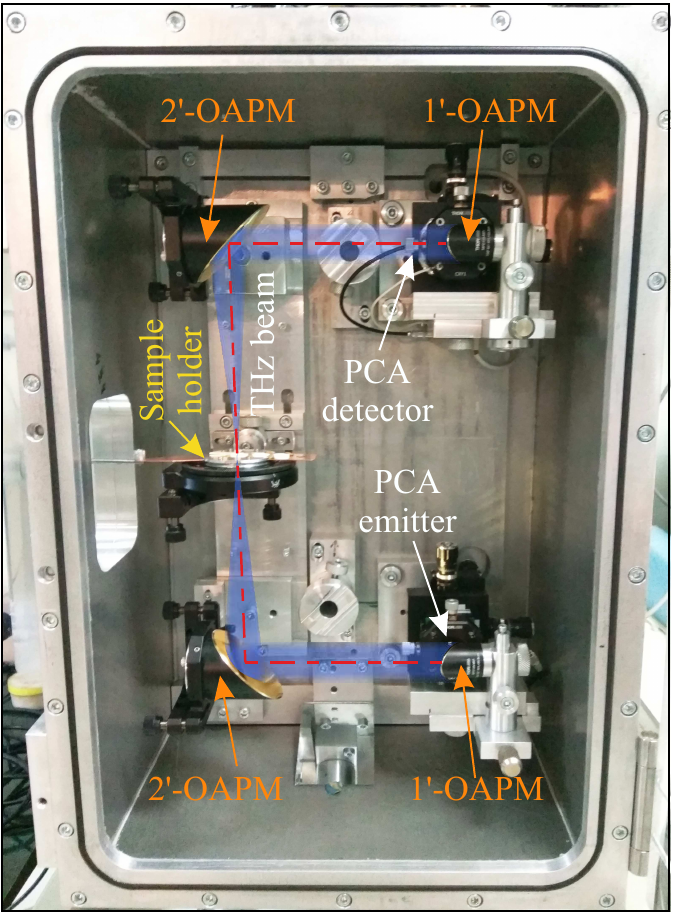}
    \caption{THz beam path of the experimental setup applied for examining the LT-GaAs PCAs performance.
    $1'$-OAPM and $2'$-OAPM stand for off-axis parabolic mirrors
    with the diameter of $1'$ and $2'$, respectively.
    The optical part located on the opposite side of the breadboard is not shown for simplicity.}
    \label{FIG:Setup}
\end{figure}

For the experimental measurements, we use an original THz pulsed spectrometer
developed in A.M. Prokhorov GPI RAS (see Fig.~\ref{FIG:Setup}).
As a source and a detector of the THz pulses,
one of the
fabricated LT-GaAs PCAs (see Tab.~\ref{TAB:Table2}),
and the commercially-available LT-GaAs
PCAs from
Fraunhofer IPM were used.
The antenna-emitter is biased by the oscillating voltage
featuring the rectangular time-domain profile, the amplitude of $10$~V, and the modulation frequency of $10$~kHz.
The current of the antenna-detector is demodulated at the same frequency in order to improve the signal-to-noise ratio.
To pump the antenna-emitter and
probe the antenna-detector,
we utilized femtosecond laser pulses of the compact all-fiber-based Toptica FemtoFErb780 laser (see Tab.~\ref{tabular:ModelPAR} for the the duration and the repetition rate of the femtosecond pulses).

The intensities of the pump and probe beams radiating the PCAs were around $10$~mW,
and the delay between the pumped and probed beams
was varied using the mechanical linear motorized delay stage (Physics Instruments LMS$60$)
featuring the maximal scanning range of 65~mm
and the
positioning
accuracy
of 0.015~$\mu$m.
In a standard configuration with the commercially available LT-GaAs antennas,
this THz pulsed spectrometer yields measurements in the spectral range of $0.1$ to $4.0$~THz
with the best spectral resolution of 0.005~THz,
and the time-domain dynamic range of 75~dB.
This setup yields measurements in vacuum and at cryogenic temperatures.
However, for examining the LT-GaAs PCA performance,
we limited ourselves to the THz measurements in non-dried room air and at room temperature.

During the experimental measurements of the fabricated LT-GaAs PCAs,
all parts of the experimental setup besides the antenna-emitter
are rigidly fixed
(the plano-convex lens,
employed for collimation of the THz beam from the antenna-emitter
and mounted separately from the LT-GaAs).
Each of the original antenna-emitters
were inserted into the experimental setup,
equally biased by the electric voltage,
and aligned in order to achieve the maximal amplitude of the THz waveform.

\subsection{Experimental Results and Their Comparison with Theory}

\begin{figure}[!t]
    \centering
    \includegraphics[width=0.8\columnwidth]{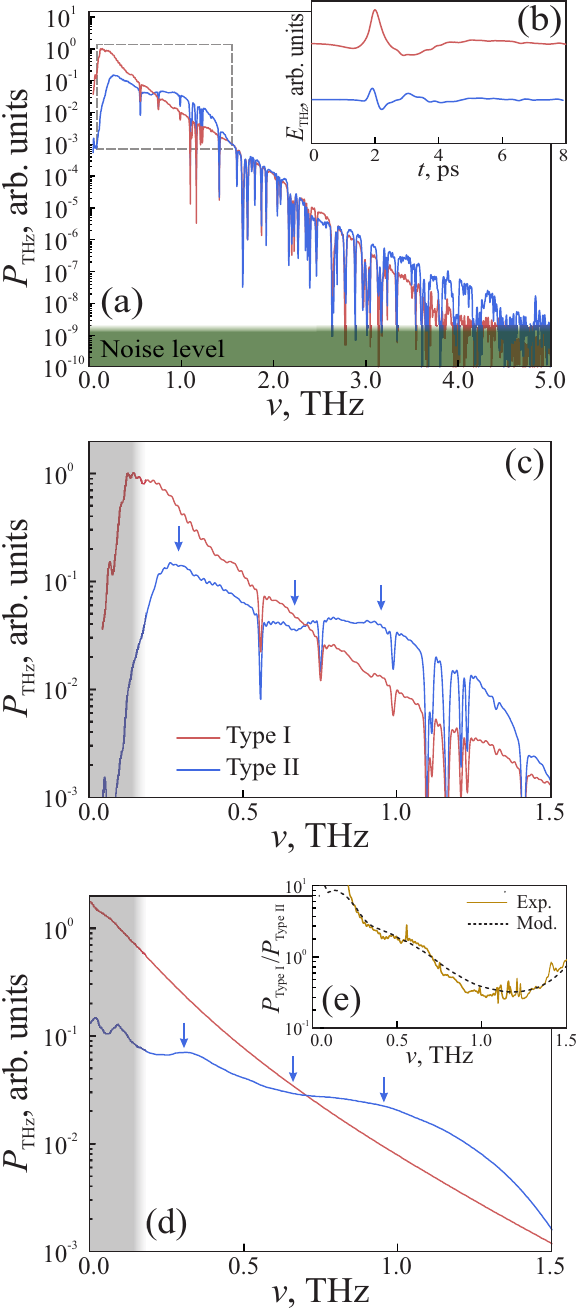}
    \caption{Experimental and simulated data on the THz pulse generation in the log-spiral PCAs with two different topology (see Tab.~\ref{TAB:Table2}):
    (a)~full-scale Fourier spectra of the fabricated PCAs with noise level;
    (b)~waveforms of the THz pulsed spectrometer;
    (c)~enlarged image of the Fourier spectra in the frequency range below 1.5 THz;
    (d)~simulated THz power spectra of the PCAs based on Figs.\ref{FIG:Ipc}--\ref{FIG:Za} at $R_\mathrm{s} \simeq 0.4$~$k \Omega$;
    (e)~ratio of the THz power spectra of PCA Type I and II, calculated using numerical and experimental data.
    In (c) and (d), arrows indicate the local maxima and minimum of the experimental and simulated THz power spectra for the PCA Type II.
    \label{FIG:Spectra}}
\end{figure}

Results of the modelling and THz measurements of the PCAs are illustrated in Fig.~\ref{FIG:Spectra}:
(a) and (c)~demonstrate the THz power spectra
$P_\mathrm{THz} \propto | \widetilde{E} ( \nu ) |^{2}$,
(where $\widetilde{E}$ is
the amplitude Fourier spectra of the waveforms);
while
(b) shows the THz waveforms $E \left( t \right)$,
generated by the fabricated PCAs and detected by the commercial one.
The observed numerous vertical lines are associated with the resonant THz absorption by water vapor along the beam path;
however, these lines do not prevent analysis and comparison of the THz power spectra of the PCAs.
During waveform detection we use the time-domain step of $50$ fs, the window size of $100$ ps (provides the frequency domain resolution of about $0.02$ THz), and the integration time of $0.1$ s for the both measured PCAs.

From Fig.~\ref{FIG:Spectra}~(b) and~(c),
we could observe a complex behaviour of the THz waveform and spectrum
generated by the PCA Type II.
Particularly, the THz pulse emitted by this antenna
is dispersed stronger and features two maxima around $2.0$--$3.5$~ps,
while for the PCA Type I we do not observe any waveform features.
Furthermore,
the PCA Type II demonstrates a modulated spectral character
with two local peaks, centered at around $0.28$ and 0.95~THz,
and a dip, centered at around $0.66$~THz (these are indicated by arrows in Fig.~\ref{FIG:Spectra}~(c) and~(d)).
In contrast, the PCA Type I shows monotonically decreasing power spectrum,
which crosses the curve of the PCA Type II at around $0.70$~THz.

The simulated power spectra of the PCAs are presented in Fig~\ref{FIG:Spectra}~(d).
One can see that PCA Type II demonstrates two local maxima and one minimum (indicated by arrows)
quite similar to the experimental curve,
and PCA Type I shows decreasing power spectra analogous to the experimental one. We note that the experimental low-frequency region (shadowed in (c), (d)) includes diffraction losses which are not considered in the numerical calculations, hence it is excluded from the comparison.

We should also point out that only a physical process of
the THz wave-emission
in PCA is considered in the numerical model, while the experimental data account for the spectral power of
emitter and the spectral sensitivity of
detector. Nevertheless, we can compare the numerical and experimental results qualitatively.
For this purpose,
we estimated ratios between the THz power spectra of the PCA Type I and Type II
for both the numerical and experimental data.
The corresponding curves are shown in Fig.~\ref{FIG:Spectra}~(e)
and demonstrate excellent agreement between the experiment and the theory.
Since the power spectra of the photocurrent in Fig.~\ref{FIG:Ipc}~(b)
has a monotonically-decreasing character
and do not possesses any resonant spectral features,
we conclude that the modulation of the THz power spectra in Fig.~\ref{FIG:Spectra}~(c) and~(d) ,
is caused by the frequency-dependent character of the antenna impedance observed in Fig.~\ref{FIG:Za},
and by changing the coupling efficiency parameter $\eta_ \mathrm{m}$ (Eq.\eqref{equation:Nm}).

The observed impact of the antenna impedance modulation on the THz pulse generation in PCA
opens an alternative way for optimizing the THz PCA performance,
in particular, for shaping the spectrum of THz pulse generation
by managing the parameters and geometry of the electrodes and of the photoconductor.
Modern semiconductor technologies
together with technologically reliable semiconductors,
such as GaAs, InGaAs and others,
form a promising fabrication and material platform
for solving the problems of the PCA optimization and the impedance matching.
In this paper, we showcase an impact of the impedance modulation on the THz pulse generation
in the particular log-spiral configurations of PCA.
The in-depth analysis and optimization of the THz pulse generation in PCAs of different type
is the subject of additional comprehensive studies. Such
studies are beyond the scope of this paper and would be considered in our future research work.

\section{Conclusions}

In this paper we predicted theoretically and justified experimentally
an  approach for shaping the spectra of the THz pulsed generation in PCA
by the frequency-dependent modulation of antenna impedance.
The results of our study demonstrated excellent correlation
between the calculated and measured THz power spectra of PCAs.
Despite considering log-spiral PCAs based on the LT-GaAs to showcase a potential of the proposed approach,
it could also be applied to other types of photoconductors
(i.e. exploiting different materials and laser pump
\cite{ApplPhysLett.71.19.2743.1997,IEEETransTerSciTech.2.6.617.2012,TechPhysLett.43.11.1020.2017,ApplPhysLett.54.24.2424.1989,ApplOpt.36.30.7853.1997,JApplPhys.116.4.043107.2014,ApplPhysExp.5.2.022601.2012,ApplPhysLett.26.3.101.1975,OptExp.18.5.4939.2010,OptExp.15.8.5120.2007})
and other antenna topology
\cite{BOOK.AntennaTheory.2005},
as well as for modelling of the continuous-wave THz generation in PCA
\cite{IEEEJQuantumElectron.41.5.717.2005}.
The proposed approach for shaping the THz pulse spectrum
can be used for a broad range of
THz technology applications
in condensed matter physics
\cite{RevModPhys.83.2.543.2011}
and material science
\cite{OptExp.16.21.17039.2008},
gas sensing
\cite{OptLett.21.24.2011.1996}
and chemistry
\cite{JPharmPharm.59.2.209.2007},
biology and medicine
\cite{TrendsBiotech.34.10.810.2016}.
Finally, we would like to stress that
the considered impedance modulation approach might be useful for optimizing the performance of the PCA-detector,
in particular, for broadening and managing the frequency-dependent sensitivity of PCA-detector
\cite{JApplPhys.109.6.061301.2011}.

\section*{Acknowledgements}

Numerical modelling and fabrication of THz PCAs
were supported by the Russian Scientific Foundation (RSF), Project \#~18-79-10195.
Experimental characterization of the THz PCAs was supported by RIEC ICRP (grant H30/A04).
The authors are also grateful to S.S.~Pushkarev for fabrication of the UV-lithography masks.





%
%

\end{document}